\documentclass[twocolumn,appendixfloats,tighten]{aastex6}

\usepackage{graphicx}	
\usepackage{amsmath}	
\usepackage{amssymb}	
\usepackage{units}
\usepackage{enumitem}
\usepackage{bm}
\usepackage{color}

\newcommand{\kms}{\, {\rm km\, s}^{-1}}

\newcommand{\be}{\begin{equation}}
\newcommand{\ee}{\end{equation}}

\def\h2{${\rm\,H_2}$}

\newcommand{\nsecure}{10}

\usepackage{color}

\begin{document}

\title{Confronting the magnetar interpretation of Fast Radio Bursts through their host galaxy demographics}

\author{Mohammadtaher Safarzadeh\altaffilmark{1,2}, J. Xavier Prochaska\altaffilmark{1,3}}
\altaffiltext{1}{Department of Astronomy and Astrophysics, University of California, Santa Cruz, CA 95064, USA 
\href{mailto:msafarza@ucsc.edu}{msafarza@ucsc.edu}}
\altaffiltext{2}{Center for Astrophysics | Harvard \& Smithsonian, 60 Garden Street, Cambridge, MA}
\altaffiltext{3}{Kavli Institute for the Physics and Mathematics of the Universe (Kavli IPMU), 5-1-5 Kashiwanoha, Kashiwa, 277-8583, Japan}
\author{Kasper E. Heintz\altaffilmark{4}}
\altaffiltext{4}{Centre for Astrophysics and Cosmology, Science Institute, University of Iceland, Dunhagi 5, 107 Reykjav\'ik, Iceland}
\author{Wen-fai Fong\altaffilmark{5}}
\altaffiltext{5}{
Center for Interdisciplinary Exploration and Research in Astrophysics (CIERA) and Department of Physics and Astronomy, Northwestern University, Evanston, IL 60208, USA}

\begin{abstract}
We explore the millisecond magnetar progenitor scenario in the context of fast radio burst (FRB) host galaxies demographics and offset distributions. Magnetars are neutron stars with strong magnetic fields on the order of $10^{15}$ G with a short decay lifetime of less than $10^4$ years. Due to their extremely short lifetimes, magnetars should follow the demographics of galaxies according to their current star-formation rate (SFR). Moreover, we hypothesize that magnetars should follow the SFR profile within galaxies, which we assume to follow an exponential profile. We construct a simple model for the host galaxies of magnetars
assuming these events track SFR in all galaxies and compare it to observed properties from a sample of \nsecure\ secure FRB hosts.  We find the distribution of observed SFRs is inconsistent with the model at $>95\%$ c.l. The offset distribution is consistent with this scenario; however, this could be due to the limited sample size and the seeing limited estimates for the effective radii of the FRB host galaxies. Despite the recent association of an FRB with a magnetar in the Milky Way, magnetars may not be the only source of FRBs in the universe, yet any other successful model must account for the demographics of the FRB host in SFR and their observed galactocentric offsets.

\end{abstract}

\section{Introduction}\label{sec:intro}
Fast radio bursts (FRBs) are brilliant ($10^{37}~K$) GHz milliseconds duration pulses \citep[see][for recent reviews]{Petroff2019,Cordes2019}. 
To explain these enigmatic millisecond-duration radio flares, a whole suite of theories have been proposed ranging from the collision of asteroids with neutron stars to magnetars \citep[see][for a compilation of the current theories]{Platts:2019ew}. 

The recent association of FRB\,200428 with a magnetar SGR\,1935+2154 in the Milky Way \citep{CHIME2020,Bochenek:2020ux}, suggests a correlation between magnetars and FRBs. Altough FRB\,200428 is two orders of magnitude fainter than the weakest extragalactic FRBs, it is still possible that this magnetar produced a burst that constitutes the faint end of the extragalactic FRB sources. Such weak FRBs would reach the detection limit of current facilities if they originated at redshifts $z>0.1$, and are therefore difficult to detect at cosmological distances. On the other hand, we could imagine two separate classes of magnetars that could be associated with FRBs. One that comes from millisecond magnetars, and one from ordinary magnetars.

If the previous history with other transient classes
holds \citep[e.g.][]{Fong:2013bc,2014ApJ...787..138L,
2020arXiv200805988S,
2016ApJ...817..144B},
studying the host galaxy properties and the physical 
offsets may be critical to constraining the likely progenitor channels of FRBs. For example, if FRBs originate from millisecond magnetars, we expect the host galaxies of such FRBs to trace the demographics of galaxies in star formation rate. 

While more than $\sim 100$ FRBs have now been detected \citep{Petroff2016}\footnote{Published online here: \url{http://frbcat.org}.}, only a dozen have been accurately localized and associated with a host galaxy \citep{Bhandari2020,Heintz2020}\footnote{See also \url{http://frbhosts.org}.}. The first host galaxy detection of an extragalactic FRB, FRB\,121102 in a starburst galaxy at $z=0.1927$ \citep{Chatterjee2017,Tendulkar2017}, showed remarkable similarities with galaxies hosting long-duration gamma-ray bursts (LGRBs) and superluminous supernovae (SLSNe), motivating a scenario where FRBs are produced by young, millisecond magnetars \citep{Metzger:2017dk}. However, later studies revealed that the host galaxy of FRB 121102 is anomalous compared to most other FRB hosts, which are typically massive galaxies with modest star-formation rates \citep{Bhandari2020}. These global properties are more consistent with the hosts of short-duration GRBs (SGRBs) and core-collapse or Type Ia SNe \citep{LiZhang2020}. The observed large physical offsets of FRBs relative to their host galaxy centers also support the latter scenario, with most bursts appearing to occur in the lower surface brightness regions of their hosts \citep{Heintz2020}. These results demonstrate that FRBs are produced in a variety of host galaxy environments. If they all originate from magnetars, this also suggests that they may be produced through a range of progenitor channels, as various channels have different expectations for host galaxy demographics \citep{Nicholl2017,Margalit2019}, or via a
single progenitor, which can accommodate a diverse range of host properties (e.g., delay time distributions) and galactocentric offsets.

The typical large offsets observed in the case of SGRBs, however, is attributed to binary neutron stars that are born with significant velocities up to hundreds of $\kms$ traveling for tens of Myr to several Gyr to reach such far distances from their host galaxies \citep{Fong:2013bc,Zevin2019}. The same logic cannot be applied to FRBs assuming that they originate from millisecond magnetars, since their large magnetic field strengths ($B\gtrsim10^{14}$ G; \citealt{Kouveliotou:1998bm}) should decay on the decay-lifetime of no more than $\tau\sim10^4$ years \citep{Colpi:2000cr,DallOsso:2012ej,Vigano:2013ij,Beniamini:2019jw}.
This short active lifetime limits the millisecond magnetars' location to their birthplace. In the context of the millisecond magnetar scenario, the observed large physical offset distribution of FRBs can, therefore, only be accounted for if there is a non-negligible probability for the millisecond magnetar to be born at their observed location. 

\begin{deluxetable*}{lccccccc}
\tablewidth{0pc}
\tablecaption{FRB host galaxy properties. \label{tab:hostprop}}
\tabletypesize{\footnotesize}
\tablehead{\colhead{FRB host}
& \colhead{$z_{\rm FRB}$}
& \colhead{$M_{\star}$}
& \colhead{SFR}
& \colhead{Offset}
& \colhead{$R_{\rm eff}$}
\\& & \colhead{($10^{9}\,M_{\odot}$)} & \colhead{($M_{\odot}$\,yr$^{-1}$)} & \colhead{(kpc)} & \colhead{(kpc)}
}
\startdata
121102 & 0.1927 & $0.14\pm 0.07$ & $0.15\pm 0.04$ & $0.6\pm 0.3$ & $0.7\pm 0.1$ \\
180916 & 0.0337 & $2.15\pm 0.33$ & $0.06\pm 0.02$ & $5.5\pm 0.1$ & $3.6\pm 0.4$ \\
180924 & 0.3212 & $13.2\pm 5.1$ & $0.88\pm 0.26$ & $3.4\pm 0.6$ & $2.7\pm 0.1$ \\
190102 & 0.2912 & $3.39\pm 1.02$ & $0.86\pm 0.26$ & $2.0\pm 2.0$ & $4.4\pm 0.5$ \\
190608 & 0.1178 & $11.6\pm 2.8$ & $0.69\pm 0.21$ & $6.6\pm 0.6$ & $2.8\pm 0.2$ \\
190611 & 0.3778 & $\sim 0.8$ & $0.27\pm 0.08$ & $11\pm 4$ & $2.1\pm 0.1$ \\
190711 & 0.5220 & $0.81\pm 0.29$ & $0.42\pm 0.12$ & $3.2\pm 2.8$ & $2.9\pm 0.2$ \\
190714 & 0.2365 & $14.9\pm 7.1$ & $0.65\pm 0.20$ & $1.9\pm 0.6$ & $3.9\pm 0.1$ \\
191001 & 0.2340 & $46.4\pm 18.8$ & $8.06\pm 2.42$ & $11\pm 1$ & $5.5\pm 0.1$ \\
200430 & 0.1600 & $1.30\pm 0.60$ & $\sim 0.2$ & $3.0\pm 1.6$ & $1.6\pm 0.5$ \\
\hline
\enddata
\tablecomments{The stellar masses $M_\star$ reported here are computed from SED modeling of the photometry for each host galaxy. The SFRs are derived based on the integrated H$\alpha$ line flux (except for the host of FRB\,200430 for which we adopt the SFR from the best-fit SED model). The offsets represent the host-burst separation and the effective radii are derived from Galfit analyses. For more detail, see \citet{Heintz2020} from which all measurements are adopted.}
\end{deluxetable*}

In this \emph{Letter} we test the hypothesis that FRBs are associated with millisecond magnetars with a simple
model for the host galaxy population of such magnetars.
While other works have tried to constrain this via direct comparisons to other astronomical transient sample \citep[e.g.][]{Bhandari2020,LiZhang2020,Heintz2020,2020arXiv200913030B}, we here provide an independent test based on theoretical expectations.
In \S~2 we describe our method to estimate the probability of accounting for the observed offset of FRBs given their short lifetime, and in \S~3 we present our results.
In \S~4 we discuss the role of natal kicks and show its irrelevance in the context of millisecond magnetar engine interpretation of FRBs, unlike that of SGRBs.
Finally, in \S~5 we discuss the implications of the offset data for the progenitors of the FRBs and conclude on our work. 

\section{Data and methods}
\label{sec:data}

In this work, we adopt the FRB host galaxy stellar population properties derived by \citet{Heintz2020}. 
For our analysis, we only include 
the \nsecure\ most secure host galaxies in their Sample~A, 
excluding FRBs\,181112, 190614D, and 190523, which have less robust host associations. The basic properties of these hosts, such as the redshift, stellar mass $M_\star$, star-formation rate (SFR), and the effective radius $R_e$ are summarized in Table~\ref{tab:hostprop}. 
We note that approximately half of the hosts exhibit 
LINER-like line-emission that may imply 
an over-estimate of the true SFR, which were primarily derived from H$\alpha$ fluxes \citep{Heintz2020}.
We further
note that the $R_e$ values were derived primarily from seeing-limited
observations, and the true measures may be systematically smaller.

Here, we also include the measured projected physical offsets of the FRBs relative to their host galaxy centers.

We then compare the observed distribution of the FRB host galaxies, in terms of their stellar mass, SFR, and their offset distribution, with theoretical expectations based on the millisecond magnetar progenitor model. To do so, we take the following steps:
\begin{enumerate}[label=\roman*.]
    \item At a given redshift, we sample the distribution from the halo mass function \citep{Tinker:2008ja}. We limit our sample to halos with dark matter halo mass $M_h>10^{10}M_{\odot}$.\footnote{We use the HMF Python package to perform sampling from their halo mass function. \url{https://ascl.net/1412.006}}
    \item We assign to each halo a stellar-mass following the stellar-mass to halo-mass relation derived in halo-abundance matching techniques \citep[e.g., ][]{Behroozi:2013fg}.
    \item To each galaxy we assign a star formation rate (SFR) based on the SFR-$M_\star$ relation of main-sequence star-forming galaxies from \citet{Whitaker:2012fs}.
    \item We apply the model assumption that the FRB host galaxies are selected according to their SFR. This is based on the assumption that since magnetars originate from massive stars, they will occur in galaxies with active star formation. Therefore, for each host, we assign a weight ($w_i$) to each galaxy $(i)$ proportional to its SFR.
    \item We probabilistically consider a fraction of galaxies to be quenched based on their stellar mass following fits to the results presented in \citet{Behroozi2019}. The quenched galaxies are then removed from the sample.
    \item We subsequently construct the cumulative distribution function (CDF) of the stellar mass distribution of all the galaxies at a given redshift. 
    \item We compare this global CDF to the corresponding CDF of the FRB hosts through a two sample KS statistics and repor the associated $p$-values.

\end{enumerate}

\section{Results}
\begin{figure*}
\includegraphics[width=\columnwidth]{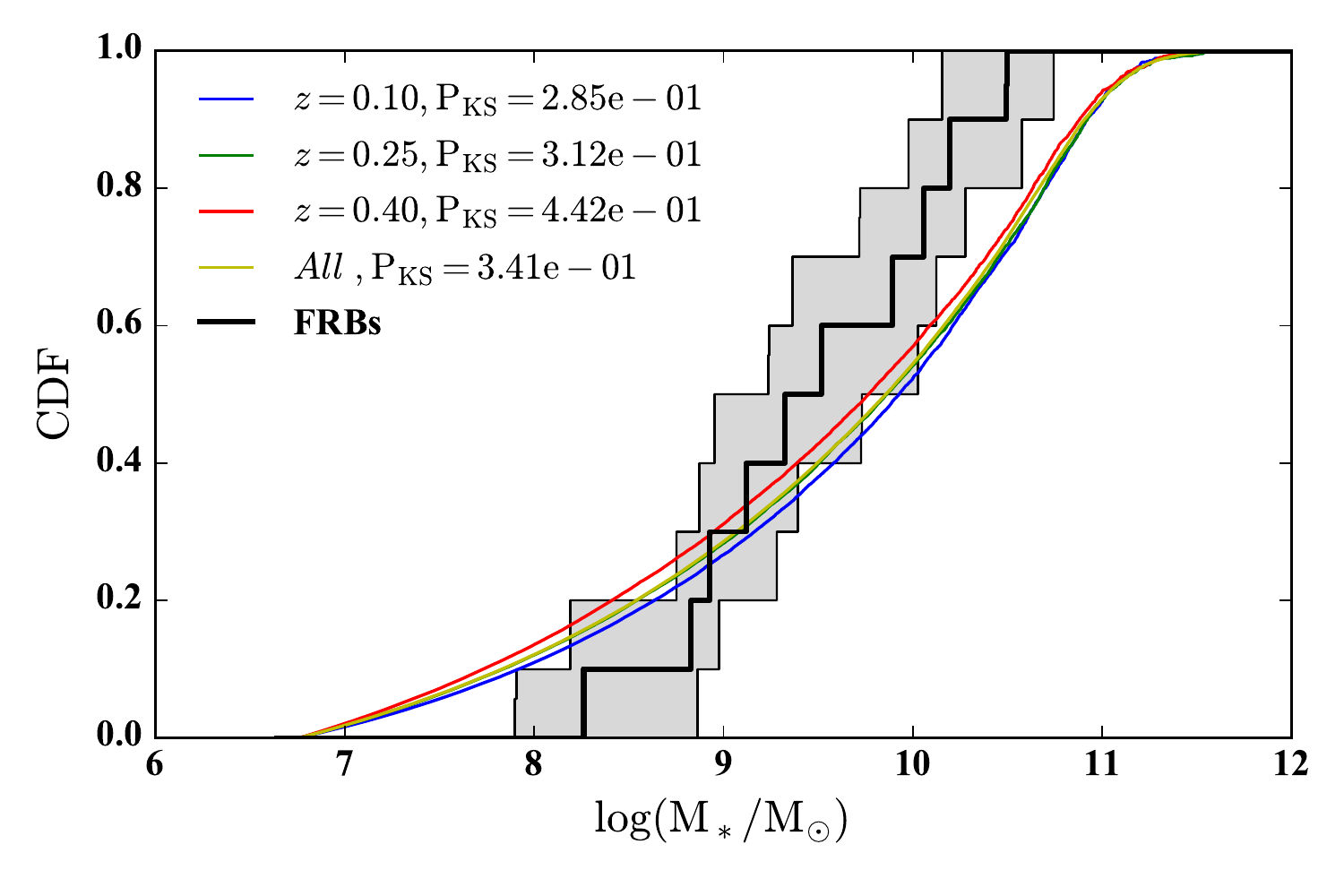}
\includegraphics[width=\columnwidth]{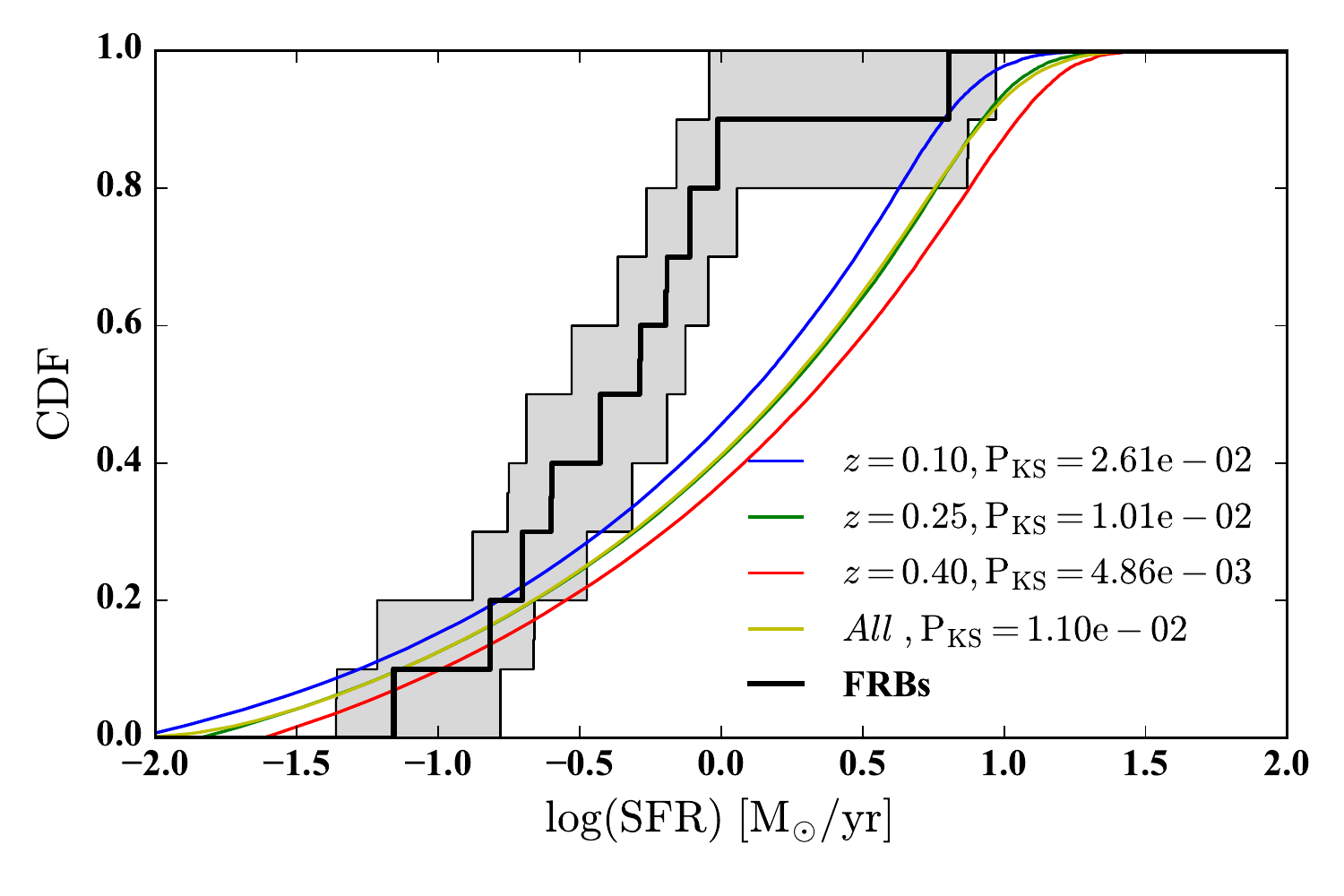}
\caption{Comparing the CDF of FRB host galaxies to the expected global distribution of galaxies at different redshift bins. \emph{Left panel}: Comparing the CDF for the stellar mass $M_\star$ distribution. The black line shows the corresponding CDF of the inferred values from
FRB hosts, while the gray region indicates the 68\% credible interval. The colored lines show the theoretical expectation of the global population of galaxies at different redshift bins.
The corresponding KS test $p$-value when compared to the median of the CDF is indicated in the legend for each line. 
\emph{Right panel:} The same shown but for SFR distribution
which is our model for the host population of magnetars as
these are expected to follow the recent star formation distribution. 
Even for $z=0.1$, this model over-predicts the observed
SFRs of the FRB host galaxies. In all panel the yellow lines are constructed by combining the individual CDF for each FRB host galaxy given the redshift, which closely follows the CDF theoretical CDF constructed based on the median redshift of our sample.
}
\label{fig:M_star_SFR_CDF}
\end{figure*}

A comparison of the inferred stellar mass and SFR distributions of the FRB hosts with the theoretical expectation of the global population of the galaxies in different redshift bins is shown in Figure~\ref{fig:M_star_SFR_CDF}. The uncertainty regions shown on the CDFs take into account both the uncertainty on each measurement and from the sample size (following \citealt{Heintz2020}, see also \citealt{Palmerio2019}). The models assume that FRB hosts track the SFR-weighted $M_\star$ and SFR distribution of field galaxies. We report the $p$-values from a two-sample KS test between the observed FRB sample in SFR and stellar mass, and their expected theoretical distribution based on the scaling relations. 
The overall distribution of $M_\star$ in FRB hosts are roughly consistent with the global CDF of the underlying galaxy population, here considered at $z\sim 0.1,$ 0.25, and 0.4 (which represents the mean and 1-$\sigma$\,c.l. of the FRB redshift distribution). 
The distribution of SFRs in FRB hosts are, however, 
inconsistent with the magnetar model at $>95\%\,$c.l.\ 
at all redshifts. A millisecond magnetar origin for most FRBs is thus difficult to reconcile with the current data.

\begin{figure}
\includegraphics[width=\columnwidth]{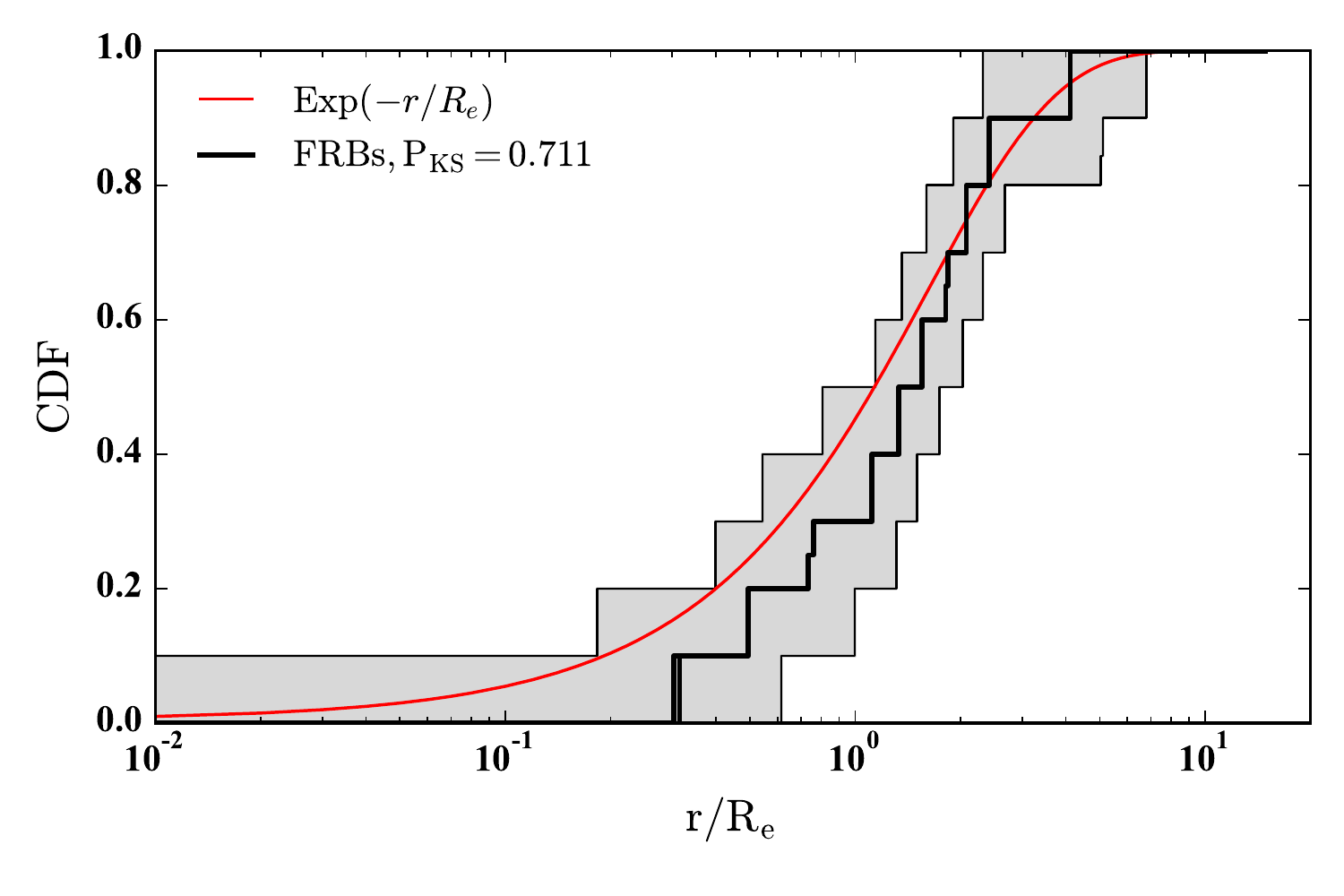}
\caption{Comparing the offset distribution of the FRBs to the expected distribution of the magnetars assuming they follow an exponential profile within galaxies. 
The red line shows the expected theoretical CDF for $r/R_e$ of the magnetars. 
The solid black line shows the distribution of the 10 FRBs. At the moment, the offset distribution is consistent with being drawn from the exponential profile. However, if future host detections result in a similar distribution to the current 10 FRB host offset distribution, the magnetar model would be incompatible with statistical significance.
}
\label{fig:r_over_Re}
\end{figure}

Next, we analyze the offset distribution of the FRBs with the local distribution of SFR
within galaxies.
For this analysis, we further assume that the SFR in these galaxies follows an exponential profile. The probability of a magnetar to be born at a projected distance $\gtrsim r$ from the host galaxy's center is computed as:
  
\be
p=\frac{\int^{\infty}_{r} r e^{-r/R_e}dr} {\int^{\infty}_{0} r e^{-r/R_e}dr}.
\ee
We construct a CDF for $r/R_e$ and compare this to the observed distribution of the host-normalized offsets $r/R_e$ of the FRBs in Figure~\ref{fig:r_over_Re}.
The red line shows the expected theoretical CDF for $r/R_e$ of magnetars in this model.
The solid black line shows the distribution of the 10 FRBs, where we arrive at a $p$-value of 0.71 when compared to the model. This indicates that with the current data, the offset distribution is consistent with being drawn from the exponential profile. 

Therefore, while the host galaxy model that we have constructed
here shows tension in terms of the global population,
the data are consistent with the simple expectation for the 
local distribution.  We caution, however, that the
$R_e$ values adopted for this analysis may be systematically too large as they were derived
almost entirely from ground-based imaging.

\section{Discussion}
\label{sec:discussion}

While there is a clear inconsistency in the expected global host demographics in SFR and the observed FRB host galaxies, we find that the projected offsets are {\it not} inconsistent with the predicted locations of magnetar birth sites. This tension may be resolved with a larger sample of well-localized FRBs, i.e.\  the consistency may primarily reflect the limited sample size. 
Sampling from the current distribution of the offset distribution and increasing the sample size to more than 50 would point to an inconsistent distribution of the FRB hosts and the assumed exponential profile. In this case, at least some FRBs may instead be produced through a ``delayed channel'', in which the delay times involved are significant compared to millisecond magnetar or massive star progenitor channels (up to several Gyr; \citealt{Nicholl2017}). Such is the case for short GRBs; indeed, their locations are not consistent with exponential disk profiles, nor the distribution of stellar mass or star formation within their hosts \citep{Fong:2013bc}. This has been interpreted as the result of a combination of natal kicks to their neutron star progenitors, and the range of expected merger timescales \citep{Paterson2020}. If FRB locations and hosts exhibit similar properties, this would indicate that some FRBs could still originate from magnetars, but via a ``delayed'' channel \citep{Margalit2019,ZhongDai2020}. However, the inferred rates of FRBs in comparison to existing delayed channel transients would also require a large fraction of FRBs to repeat \citep{Ravi19,Wang20}. In future work, we will examine the required delay times for such a scenario to make the magnetar interpretation still be viable.

Our current model could be further improved from different aspects: 
 1) allowing for separation of the repeaters and non-repeaters, which we have avoided in this work due to the limited sample size; 
 2) assuming other internal galaxy properties (e.g., \ metallicity) influence the progenitor model; 
 3) comparing to the observed SFR distribution in FRB
 hosts instead of adopting a simple exponential profile;
 3) investigating the uncertainties and systematics
 related to the SFR estimates, especially between 
 those adopted for the FRB host galaxies \citep[derived
 primarily from slit spectroscopy of H$\alpha$;][]{Heintz2020}  
 and those from the literature;
 and
 4) examination of other observed galaxy properties
 that may be expected to trace millisecond magnetars
 (e.g.,\ age). 
These will be the focus of future works, especially
as the FRB host galaxy sample will inevitably increase with the onset of new or upgraded discovery experiments.

\section{Conclusions and outlook}

Here, we have examined the demographics and galactocentric offsets of FRBs based on the global properties of their hosts to test the assumption that the majority of these bursts are produced by millisecond magnetars. 

Based on the $M_\star$ and SFR distribution of the most recent sample of 10 securely identified FRB host galaxies, we here showed that there is an apparent tension between the observed distributions and theoretical expectations of galaxies hosting active magnetars. Due to their short decay-lifetimes, active magnetars would be expected to follow star formation, which is inconsistent at $>95\%$\,c.l.\ with the observed distributions of FRB hosts. Alternatively, we found that the host-normalized offset distribution of the current sample of FRB hosts cannot rule out the scenario where FRBs originate from magnetars. This consistency could be due to limited sample size at hand.
While the current cadence is still relatively low, of the order $\sim 5$ arcsecond-localized FRBs per year (required to robustly identify their associated hosts), this detection rate is expected to accelerate within the next few fears. Once a sample size of $\sim 100$ FRB hosts have been reached, this will be sufficient to test more sophisticated models for the galaxy populations of assumed progenitor models.

\section*{Acknowledgements}
We would like to thank Jesse Palmerio for sharing his code used to produce the CDFs and related uncertainties shown in the paper.
We thank Nicol\'as Tejos Salgado for comments on the earlier version of this work.
The Fast and Fortunate for FRB
Follow-up team, acknowledge support from 
NSF grants AST-1911140 and AST-1910471. 
K.E.H. acknowledges support by a Project Grant (162948--051) from The Icelandic Research Fund. 
W.F. acknowledges support by the National Science Foundation under grant No. AST-1814782.
 
\bibliographystyle{yahapj}
\bibliography{wf_refs}

\begin{thebibliography}{}
\providecommand\natexlab[1]{#1}
\providecommand\JournalTitle[1]{#1}

\bibitem[{{Behroozi} {et~al.}(2019){Behroozi}, {Wechsler}, {Hearin}, \&
  {Conroy}}]{Behroozi2019}
{Behroozi}, P., {Wechsler}, R.~H., {Hearin}, A.~P., \& {Conroy}, C. 2019,
  \href{http://dx.doi.org/10.1093/mnras/stz1182}{\JournalTitle{\mnras}, 488,
  3143}

\bibitem[{{Behroozi} {et~al.}(2013){Behroozi}, {Wechsler}, \&
  {Conroy}}]{Behroozi:2013fg}
{Behroozi}, P.~S., {Wechsler}, R.~H., \& {Conroy}, C. 2013,
  \href{http://dx.doi.org/10.1088/0004-637X/770/1/57}{\JournalTitle{\apj}, 770,
  57}

\bibitem[{{Beniamini} {et~al.}(2019){Beniamini}, {Hotokezaka}, {van der Horst},
  \& {Kouveliotou}}]{Beniamini:2019jw}
{Beniamini}, P., {Hotokezaka}, K., {van der Horst}, A., \& {Kouveliotou}, C.
  2019, \href{http://dx.doi.org/10.1093/mnras/stz1391}{\JournalTitle{\mnras},
  487, 1426}

\bibitem[{{Bhandari} {et~al.}(2020){Bhandari}, {Sadler}, {Prochaska}, {Simha},
  {Ryder}, {Marnoch}, {Bannister}, {Macquart}, {Flynn}, {Shannon}, {Tejos},
  {Corro-Guerra}, {Day}, {Deller}, {Ekers}, {Lopez}, {Mahony}, {Nu{\~n}ez}, \&
  {Phillips}}]{Bhandari2020}
{Bhandari}, S., {Sadler}, E.~M., {Prochaska}, J.~X., {et~al.} 2020,
  \href{http://dx.doi.org/10.3847/2041-8213/ab672e}{\JournalTitle{\apjl}, 895,
  L37}

\bibitem[{{Blanchard} {et~al.}(2016){Blanchard}, {Berger}, \&
  {Fong}}]{2016ApJ...817..144B}
{Blanchard}, P.~K., {Berger}, E., \& {Fong}, W.-f. 2016,
  \href{http://dx.doi.org/10.3847/0004-637X/817/2/144}{\JournalTitle{\apj},
  817, 144}

\bibitem[{{Bochenek} {et~al.}(2020{\natexlab{a}}){Bochenek}, {Ravi}, {Belov},
  {Hallinan}, {Kocz}, {Kulkarni}, \& {McKenna}}]{Bochenek:2020ux}
{Bochenek}, C.~D., {Ravi}, V., {Belov}, K.~V., {et~al.} 2020{\natexlab{a}},
  \JournalTitle{arXiv e-prints}, arXiv:2005.10828

\bibitem[{{Bochenek} {et~al.}(2020{\natexlab{b}}){Bochenek}, {Ravi}, \&
  {Dong}}]{2020arXiv200913030B}
{Bochenek}, C.~D., {Ravi}, V., \& {Dong}, D. 2020{\natexlab{b}},
  \JournalTitle{arXiv e-prints}, arXiv:2009.13030

\bibitem[{{Chatterjee} {et~al.}(2017){Chatterjee}, {Law}, {Wharton},
  {Burke-Spolaor}, {Hessels}, {Bower}, {Cordes}, {Tendulkar}, {Bassa},
  {Demorest}, {Butler}, {Seymour}, {Scholz}, {Abruzzo}, {Bogdanov}, {Kaspi},
  {Keimpema}, {Lazio}, {Marcote}, {McLaughlin}, {Paragi}, {Ransom}, {Rupen},
  {Spitler}, \& {van Langevelde}}]{Chatterjee2017}
{Chatterjee}, S., {Law}, C.~J., {Wharton}, R.~S., {et~al.} 2017,
  \href{http://dx.doi.org/10.1038/nature20797}{\JournalTitle{\nat}, 541, 58}

\bibitem[{{Colpi} {et~al.}(2000){Colpi}, {Geppert}, \& {Page}}]{Colpi:2000cr}
{Colpi}, M., {Geppert}, U., \& {Page}, D. 2000,
  \href{http://dx.doi.org/10.1086/312448}{\JournalTitle{\apjl}, 529, L29}

\bibitem[{{Cordes} \& {Chatterjee}(2019)}]{Cordes2019}
{Cordes}, J.~M., \& {Chatterjee}, S. 2019,
  \href{http://dx.doi.org/10.1146/annurev-astro-091918-104501}{\JournalTitle{\araa},
  57, 417}

\bibitem[{{Dall'Osso} {et~al.}(2012){Dall'Osso}, {Granot}, \&
  {Piran}}]{DallOsso:2012ej}
{Dall'Osso}, S., {Granot}, J., \& {Piran}, T. 2012,
  \href{http://dx.doi.org/10.1111/j.1365-2966.2012.20612.x}{\JournalTitle{\mnras},
  422, 2878}

\bibitem[{{Fong} \& {Berger}(2013)}]{Fong:2013bc}
{Fong}, W., \& {Berger}, E. 2013,
  \href{http://dx.doi.org/10.1088/0004-637X/776/1/18}{\JournalTitle{\apj}, 776,
  18}

\bibitem[{Heintz {et~al.}(2020)Heintz, Prochaska, Simha, Platts, fai Fong,
  Tejos, Ryder, Aggerwal, Bhandari, Day, Deller, Kilpatrick, Law, Macquart,
  Mannings, Marnoch, Sadler, \& Shannon}]{Heintz2020}
Heintz, K.~E., Prochaska, J.~X., Simha, S., {et~al.} 2020, Host Galaxy
  Properties and Offset Distributions of Fast Radio Bursts: Implications for
  their Progenitors, \href{http://arxiv.org/abs/2009.10747}{{\sffamily
  arXiv:2009.10747 [astro-ph.GA]}}

\bibitem[{{Kouveliotou} {et~al.}(1998){Kouveliotou}, {Dieters}, {Strohmayer},
  {van Paradijs}, {Fishman}, {Meegan}, {Hurley}, {Kommers}, {Smith}, {Frail},
  \& {Murakami}}]{Kouveliotou:1998bm}
{Kouveliotou}, C., {Dieters}, S., {Strohmayer}, T., {et~al.} 1998,
  \href{http://dx.doi.org/10.1038/30410}{\JournalTitle{\nat}, 393, 235}

\bibitem[{{Li} \& {Zhang}(2020)}]{LiZhang2020}
{Li}, Y., \& {Zhang}, B. 2020,
  \href{http://dx.doi.org/10.3847/2041-8213/aba907}{\JournalTitle{\apjl}, 899,
  L6}

\bibitem[{{Lunnan} {et~al.}(2014){Lunnan}, {Chornock}, {Berger}, {Laskar},
  {Fong}, {Rest}, {Sanders}, {Challis}, {Drout}, {Foley}, {Huber}, {Kirshner},
  {Leibler}, {Marion}, {McCrum}, {Milisavljevic}, {Narayan}, {Scolnic},
  {Smartt}, {Smith}, {Soderberg}, {Tonry}, {Burgett}, {Chambers}, {Flewelling},
  {Hodapp}, {Kaiser}, {Magnier}, {Price}, \& {Wainscoat}}]{2014ApJ...787..138L}
{Lunnan}, R., {Chornock}, R., {Berger}, E., {et~al.} 2014,
  \href{http://dx.doi.org/10.1088/0004-637X/787/2/138}{\JournalTitle{\apj},
  787, 138}

\bibitem[{{Margalit} {et~al.}(2019){Margalit}, {Berger}, \&
  {Metzger}}]{Margalit2019}
{Margalit}, B., {Berger}, E., \& {Metzger}, B.~D. 2019,
  \href{http://dx.doi.org/10.3847/1538-4357/ab4c31}{\JournalTitle{\apj}, 886,
  110}

\bibitem[{{Metzger} {et~al.}(2017){Metzger}, {Berger}, \&
  {Margalit}}]{Metzger:2017dk}
{Metzger}, B.~D., {Berger}, E., \& {Margalit}, B. 2017,
  \href{http://dx.doi.org/10.3847/1538-4357/aa633d}{\JournalTitle{\apj}, 841,
  14}

\bibitem[{{Nicholl} {et~al.}(2017){Nicholl}, {Williams}, {Berger}, {Villar},
  {Alexander}, {Eftekhari}, \& {Metzger}}]{Nicholl2017}
{Nicholl}, M., {Williams}, P.~K.~G., {Berger}, E., {et~al.} 2017,
  \href{http://dx.doi.org/10.3847/1538-4357/aa794d}{\JournalTitle{\apj}, 843,
  84}

\bibitem[{{Palmerio} {et~al.}(2019){Palmerio}, {Vergani}, {Salvaterra}, {Sand
  ers}, {Japelj}, {Vidal-Garc{\'\i}a}, {D'Avanzo}, {Corre}, {Perley},
  {Shapley}, {Boissier}, {Greiner}, {Le Floc'h}, \& {Wiseman}}]{Palmerio2019}
{Palmerio}, J.~T., {Vergani}, S.~D., {Salvaterra}, R., {et~al.} 2019,
  \href{http://dx.doi.org/10.1051/0004-6361/201834179}{\JournalTitle{\aap},
  623, A26}

\bibitem[{{Paterson} {et~al.}(2020){Paterson}, {Fong}, {Nugent}, {Escorial},
  {Leja}, {Laskar}, {Chornock}, {Miller}, {Scharw{\"a}chter}, {Cenko},
  {Perley}, {Tanvir}, {Levan}, {Cucchiara}, {Cobb}, {De}, {Berger}, {Terreran},
  {Alexander}, {Nicholl}, {Blanchard}, \& {Cornish}}]{Paterson2020}
{Paterson}, K., {Fong}, W., {Nugent}, A., {et~al.} 2020,
  \href{http://dx.doi.org/10.3847/2041-8213/aba4b0}{\JournalTitle{\apjl}, 898,
  L32}

\bibitem[{{Petroff} {et~al.}(2019){Petroff}, {Hessels}, \&
  {Lorimer}}]{Petroff2019}
{Petroff}, E., {Hessels}, J.~W.~T., \& {Lorimer}, D.~R. 2019,
  \href{http://dx.doi.org/10.1007/s00159-019-0116-6}{\JournalTitle{\aapr}, 27,
  4}

\bibitem[{{Petroff} {et~al.}(2016){Petroff}, {Barr}, {Jameson}, {Keane},
  {Bailes}, {Kramer}, {Morello}, {Tabbara}, \& {van Straten}}]{Petroff2016}
{Petroff}, E., {Barr}, E.~D., {Jameson}, A., {et~al.} 2016,
  \href{http://dx.doi.org/10.1017/pasa.2016.35}{\JournalTitle{\pasa}, 33, e045}

\bibitem[{{Platts} {et~al.}(2019){Platts}, {Weltman}, {Walters}, {Tendulkar},
  {Gordin}, \& {Kandhai}}]{Platts:2019ew}
{Platts}, E., {Weltman}, A., {Walters}, A., {et~al.} 2019,
  \href{http://dx.doi.org/10.1016/j.physrep.2019.06.003}{\JournalTitle{\physrep},
  821, 1}

\bibitem[{{Ravi} {et~al.}(2019){Ravi}, {Catha}, {D'Addario}, {Djorgovski},
  {Hallinan}, {Hobbs}, {Kocz}, {Kulkarni}, {Shi}, {Vedantham}, {Weinreb}, \&
  {Woody}}]{Ravi19}
{Ravi}, V., {Catha}, M., {D'Addario}, L., {et~al.} 2019,
  \href{http://dx.doi.org/10.1038/s41586-019-1389-7}{\JournalTitle{\nat}, 572,
  352}

\bibitem[{{Schulze} {et~al.}(2020){Schulze}, {Yaron}, {Sollerman}, {Leloudas},
  {Gal}, {Wright}, {Lunnan}, {Gal-Yam}, {Ofek}, {Perley}, {Filippenko},
  {Kasliwal}, {Kulkarni}, {Nugent}, {Quimby}, {Sullivan}, {Linn Strothjohann},
  {Arcavi}, {Ben-Ami}, {Bianco}, {Bloom}, {De}, {Fraser}, {Fremling}, {Horesh},
  {Johansson}, {Kelly}, {Knezevic}, {Maguire}, {Nyholm}, {Semeli
  Papadogiannakis}, {Petrushevska}, {Rubin}, {Yan}, {Yang}, {Adams}, {Bufano},
  {Clubb}, {Foley}, {Green}, {Harmanen}, {Ho}, {Hook}, {Hosseinzadeh},
  {Howell}, {Kong}, {Kotak}, {Matheson}, {McCully}, {Milisavljevic}, {Pan},
  {Poznanski}, {Shivvers}, \& {van Velzen}}]{2020arXiv200805988S}
{Schulze}, S., {Yaron}, O., {Sollerman}, J., {et~al.} 2020, \JournalTitle{arXiv
  e-prints}, arXiv:2008.05988

\bibitem[{{Tendulkar} {et~al.}(2017){Tendulkar}, {Bassa}, {Cordes}, {Bower},
  {Law}, {Chatterjee}, {Adams}, {Bogdanov}, {Burke-Spolaor}, {Butler},
  {Demorest}, {Hessels}, {Kaspi}, {Lazio}, {Maddox}, {Marcote}, {McLaughlin},
  {Paragi}, {Ransom}, {Scholz}, {Seymour}, {Spitler}, {van Langevelde}, \&
  {Wharton}}]{Tendulkar2017}
{Tendulkar}, S.~P., {Bassa}, C.~G., {Cordes}, J.~M., {et~al.} 2017,
  \href{http://dx.doi.org/10.3847/2041-8213/834/2/L7}{\JournalTitle{\apjl},
  834, L7}

\bibitem[{{The CHIME/FRB Collaboration} {et~al.}(2020){The CHIME/FRB
  Collaboration}, {:}, {Andersen}, {Band ura}, {Bhardwaj}, {Bij}, {Boyce},
  {Boyle}, {Brar}, {Cassanelli}, {Chawla}, {Chen}, {Cliche}, {Cook},
  {Cubranic}, {Curtin}, {Denman}, {Dobbs}, {Dong}, {Fandino}, {Fonseca},
  {Gaensler}, {Giri}, {Good}, {Halpern}, {Hill}, {Hinshaw}, {H{\"o}fer},
  {Josephy}, {Kania}, {Kaspi}, {Landecker}, {Leung}, {Li}, {Lin}, {Masui},
  {Mckinven}, {Mena-Parra}, {Merryfield}, {Meyers}, {Michilli}, {Milutinovic},
  {Mirhosseini}, {M{\"u}nchmeyer}, {Naidu}, {Newburgh}, {Ng}, {Patel}, {Pen},
  {Pinsonneault-Marotte}, {Pleunis}, {Quine}, {Rafiei-Ravandi}, {Rahman},
  {Ransom}, {Renard}, {Sanghavi}, {Scholz}, {Shaw}, {Shin}, {Siegel}, {Singh},
  {Smegal}, {Smith}, {Stairs}, {Tan}, {Tendulkar}, {Tretyakov}, {Vanderlinde},
  {Wang}, {Wulf}, \& {Zwaniga}}]{CHIME2020}
{The CHIME/FRB Collaboration}, {:}, {Andersen}, B.~C., {et~al.} 2020,
  \JournalTitle{arXiv e-prints}, arXiv:2005.10324

\bibitem[{{Tinker} {et~al.}(2008){Tinker}, {Kravtsov}, {Klypin}, {Abazajian},
  {Warren}, {Yepes}, {Gottl{\"o}ber}, \& {Holz}}]{Tinker:2008ja}
{Tinker}, J., {Kravtsov}, A.~V., {Klypin}, A., {et~al.} 2008,
  \href{http://dx.doi.org/10.1086/591439}{\JournalTitle{\apj}, 688, 709}

\bibitem[{{Vigan{\`o}} {et~al.}(2013){Vigan{\`o}}, {Rea}, {Pons}, {Perna},
  {Aguilera}, \& {Miralles}}]{Vigano:2013ij}
{Vigan{\`o}}, D., {Rea}, N., {Pons}, J.~A., {et~al.} 2013,
  \href{http://dx.doi.org/10.1093/mnras/stt1008}{\JournalTitle{\mnras}, 434,
  123}

\bibitem[{{Wang} {et~al.}(2020){Wang}, {Wang}, {Yang}, {Yu}, {Zuo}, \&
  {Dai}}]{Wang20}
{Wang}, F.~Y., {Wang}, Y.~Y., {Yang}, Y.-P., {et~al.} 2020,
  \href{http://dx.doi.org/10.3847/1538-4357/ab74d0}{\JournalTitle{\apj}, 891,
  72}

\bibitem[{{Whitaker} {et~al.}(2012){Whitaker}, {van Dokkum}, {Brammer}, \&
  {Franx}}]{Whitaker:2012fs}
{Whitaker}, K.~E., {van Dokkum}, P.~G., {Brammer}, G., \& {Franx}, M. 2012,
  \href{http://dx.doi.org/10.1088/2041-8205/754/2/L29}{\JournalTitle{\apjl},
  754, L29}

\bibitem[{{Zevin} {et~al.}(2019){Zevin}, {Kelley}, {Nugent}, {Fong}, {Berry},
  \& {Kalogera}}]{Zevin2019}
{Zevin}, M., {Kelley}, L.~Z., {Nugent}, A., {et~al.} 2019, \JournalTitle{arXiv
  e-prints}, arXiv:1910.03598

\bibitem[{{Zhong} \& {Dai}(2020)}]{ZhongDai2020}
{Zhong}, S.-Q., \& {Dai}, Z.-G. 2020,
  \href{http://dx.doi.org/10.3847/1538-4357/ab7bdf}{\JournalTitle{\apj}, 893,
  9}

\end{thebibliography}
\end{document}